\documentclass[aps,prl, showpacs,twocolumn]{revtex4}
\usepackage{amssymb}
\usepackage{amsmath}
\usepackage{graphicx}
\usepackage{epsfig}

\begin{document}

\title{Characterizing entanglement by momentum-jump \\
in the frustrated Heisenberg ring at quantum phase transition}
\author{ Xiao-Feng Qian $^{1}$, Tao Shi$^{1}$, Ying Li$^{1}$, Zhi Song$^{1,a}$ and
Chang-Pu Sun$^{1,2,a,b}$ } \affiliation{$^{1}$Department of
Physics, Nankai University, Tianjin 300071, China}
\affiliation{$^{2}$ Institute of Theoretical Physics, Chinese
Academy of Sciences, Beijing, 100080, China}

\begin{abstract}
We study the pairwise concurrences, a measure of entanglement, of
the ground states for the frustrated Heisenberg ring to explore
the relation between entanglement and quantum phase transition
associated with the momentum jump. The groundstate concurrences
between any two sites are obtained analytically and numerically.
It shows that the summation of all possible pairwise concurrences
is an appropriate candidate to depict the phase transition. We
also investigate the role that the momentum takes in the jump of
concurrence at the critical points. We find that an abrupt
momentum change rusults in the maximal concurrence difference of
two degenerate ground states.
\end{abstract}

\maketitle

\section{I. Introduction}

A quantum system changes its groundstate properties in a
fundamental way when quantum phase transitions (QPTs) occur at
absolute zero \cite{Sachdev}, which are induced by the change of
external parameters or coupling constants and are driven by
quantum fluctuations. In a finite system the discontinuity of the
groundstate energy is often used to characterize the occurrence of
QPTs for there is an energy-level crossing at the critical point.
We will show, in this paper, that such kind of phase transitions
are usually accompanied by the change of symmetry characterized by
conservative quantities such as the momentum which is the
generator of translation. Our investigations will relate to the
quantum entanglement measured by the total concurrences
\cite{chung}.

Actually, quantum critical points are governed by a diverging
correlation length and there exists a very close relation between
quantum correlation and quantum entanglement, which is known as
the resource that enables quantum computation and communication.
So exploring the role of entanglement in a phase transition has
attracted great attention from both the
communities of quantum computation and quantum statistics \cite%
{Osterloh:02,Osborne:02,JVidal:04a,Huang:04,
Vidal:03,Verstraete:04a,Barnum:04,Bose:02,Alcaraz:03,Gu:03,Lambert:04,LAWu:04}%
. People have connected the theory of critical phenomena with
quantum information by exploring the entangling resources of a
system close to its quantum critical point. They demonstrate, for
a class of magnetic systems or the interacting quantum lattice
spin systems at zero temperature, that entanglement shows scaling
behavior in the vicinity of the transition point where the level
crossing occurs at degenerate groundstates \cite{Tian}.

From these studies we observe that the degeneracy of groundstates
at the critical point may result in the uncertainty of
entanglement. This fact means the discontinuity of concurrence in
the vicinity of the phase transition point. Such sudden change of
the concurrence as the variation of an external parameter or
coupling constant is vividly called the jump of concurrence.
However, the groundstate energy-level crossing may not always
result in a jump of a certain type of concurrence, such as the NNN
or other pairwise concurrences. On the other hand, this kind of
QPTs must be accompanied by a change of a certain conservative
quantity, such as the momentum or the macroscopic magnetization.
This observation may provide us a new way to find an appropriate
definition of concurrence that just characterize the property of
the quantum spin systems at the critical point.

In this paper, we consider a frustrated Heisenberg ring system
which contains rich phases in the ground state. To reveal the
connection between the concurrence behavior and the symmetries of
the separated phases around the critical point, we study the
difference between the concurrences of two degenerate ground
states at the critical points. We find that there does exist such
a discontinuity of a conservative physical quantity--momentum (the
generator of translation) which results in the maximization of
entanglement difference of the two degenerate ground states.

The paper is organized as follows. In Sec. II, numerical result
for the concurrence behavior of a frustrated Heisenberg ring
around the critical points is given. It shows that a single type
of concurrence is not sufficient to depict a QPT, while the
summation of all types of concurrence may be. In Sec. III, the
exact results are employed to explain our observation concluded in
Sec. II. In Sec. IV, the role that another conservative quantity -
momentum plays in the concurrence jump is studied. The summary and
some discussions are given in Sec. V.

\section{II. Concurrence Jumps in QPTs}

We start from a 1D frustrated spin-$\frac{1}{2}$ Heisenberg model
with periodic boundary conditions, which belongs to the
Majumdar-Ghosh (MG) families of models \cite{MG} (see Fig. 1).

\begin{figure}[tbp]
\includegraphics[bb=65 355 525 527, width=7 cm, clip]{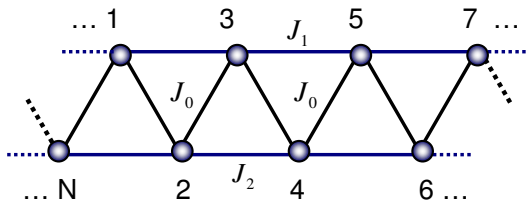}
\caption{\textit{The schematic structure of a frustrated
spin-$\frac{1}{2}$ Heisenberg ring. $J_{0}$ denotes the nearest
neighbor coupling constant, while $J_{1}$ and $J_{2}$ denote the
next nearest neighbor coupling constants between even and odd
number of sites respectively. In this paper we only concern the
simple case $J_{1}=J_{2}=J$.}}
\end{figure}
The Hamiltonian reads%
\begin{equation}
H=J_{0}\sum_{i=1}^{N}\mathbf{S}_{i}\cdot \mathbf{S}_{i+1}+J\sum_{i=1}^{N}%
\mathbf{S}_{i}\cdot \mathbf{S}_{i+2},
\end{equation}%
where $N$ is even denoting the total number of the lattice sites, $\mathbf{S}%
_{i}$ is the spin operator at $i$th site, and $J_{0}$ $(J)$ is the
strength of the NN (NNN) exchange interaction. For the periodic
boundary conditions, we have $\mathbf{S}_{N+1}=\mathbf{S}_{1}$.
For arbitrary $J_{0}$ and $J$, the groundstate energy of
Hamiltonian (1) can be formally written as the function of the
parameters $J_{0}$ and $J$
\begin{equation}
E_{g}(J_{0},J)=J_{0}h_{0}+Jh,
\end{equation}%
in terms of the NN correlation function
\begin{equation}
h_{0}=\sum_{i}\left\langle g\right\vert \mathbf{S}_{i}\cdot \mathbf{S}%
_{i+1}\left\vert g\right\rangle
\end{equation}%
and the NNN correlation function
\begin{equation}
h=\sum_{i}\left\langle g\right\vert \mathbf{S}_{i}\cdot \mathbf{S}%
_{i+2}\left\vert g\right\rangle
\end{equation}%
with respect to the ground state $\left\vert g\right\rangle $.

Now we consider the behavior of groundstate energy as the
variation of parameters $J$ and $J_{0}$. For an infinite system,
the energy-level crossing should induce the discontinuity of the
derivative of groundstate energy
\begin{equation}
\frac{\partial E_{g}}{\partial J_{0}}=\left\langle g\right\vert \frac{%
\partial H}{\partial J_{0}}\left\vert g\right\rangle =h_{0},  \label{dd}
\end{equation}%
which is just the NN correlation function. Since there exist an
algebraic
relationship between pairwise concurrence and correlation function \cite%
{Wang}
\begin{equation}
C_{ij}=\frac{1}{2}\max \left\{ 0,\left\vert
G_{xx}+G_{yy}\right\vert -G_{zz}-1\right\} ,
\end{equation}%
where $G_{\alpha \alpha }=\left\langle g\right\vert
\sigma_{i}^{\alpha }\sigma_{j}^{\alpha }\left\vert g\right\rangle
$ ($\alpha =x,y,z$) are correlation functions, then according the
Eq. (\ref{dd}) the energy-level crossing will lead to the
discontinuity of the NN concurrence. On the other hand, one can
also establish a similar relation between NNN correlation function
$h$ and the discontinuity of $\partial E_{g}/\partial J,$
\begin{equation}
\frac{\partial E_{g}}{\partial J}=\left\langle g\right\vert \frac{\partial H%
}{\partial J}\left\vert g\right\rangle =h.
\end{equation}%
Notice that, for nonzero $J_{0}$, if the correlation $h$ has a
jump due to the discontinuity of $\partial E_{g}/\partial J$, the
other correlation function $h_{0}$ must experience a jump at the
same point $J=J_{c}$. Actually, at the energy-level crossing point
the ground states are degenerates, i.e.
\begin{equation}
J_{0}h_{0}+Jh=J_{0}h_{0}^{\prime }+Jh^{\prime },
\end{equation}%
where
\begin{eqnarray}
h_{0}^{\prime } &=&\sum_{i}\left\langle g^{\prime }\right\vert \mathbf{S}%
_{i}\cdot \mathbf{S}_{i+1}\left\vert g^{\prime }\right\rangle , \\
h^{\prime } &=&\sum_{i}\left\langle g^{\prime }\right\vert \mathbf{S}%
_{i}\cdot \mathbf{S}_{i+2}\left\vert g^{\prime }\right\rangle
\notag
\end{eqnarray}%
are the corresponding correlations with respect to another ground state $%
\left\vert g^{\prime }\right\rangle $. No doubt, both the NN and
NNN
correlation functions must be discontinuous or have jumps at critical point $%
J=J_{c}$. But the jump of one type of correlation may not
necessarily induce the jump of its corresponding concurrence. A
natural question is that, which type of concurrence, NN or NNN,
plays a major role in depicting the QPTs at the critical points?
To answer this question, we investigate the frustrated
spin-$\frac{1}{2}$ Heisenberg ring numerically and analytically.
In Fig. 2, the eigen energies of ground and first excited state
are plotted for the systems with size $N=6,8,10$ and $12$. $A$ and
$B$ denote the two energy-level crossing points.

\begin{figure}[tbp]
\includegraphics[bb=40 230 570 760, width=7 cm, clip]{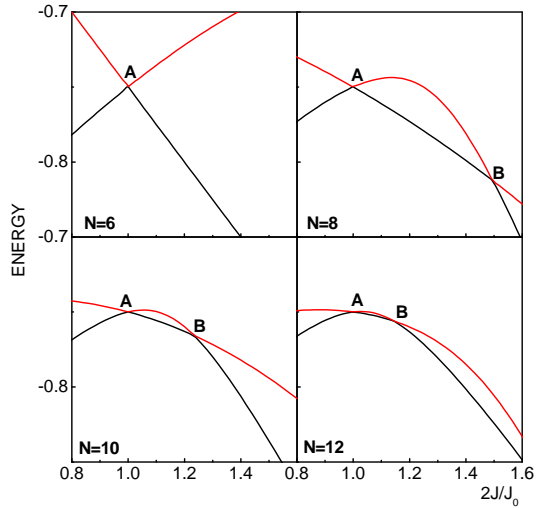}
\caption{\textit{The ground and first excited eigen energies for
the systems of the size $N=6,8,10$ and $12$. $A$ and $B$ denote
the energy level crossing points where the quantum phase
transitions occur. The critical point $A$ is always at
$J=J_{c}=J_{0}/2$, while the position of point $B$ depends on the
size of system.}}
\end{figure}

The reduced density matrix for two spins located at sites $i$ and $j$ \cite%
{entangled rings} has the form
\begin{equation}
\rho _{ij}=\left(
\begin{array}{cccc}
v_{ij} & 0 & 0 & 0 \\
0 & w_{ij} & z_{ij} & 0 \\
0 & z_{ij} & w_{ij} & 0 \\
0 & 0 & 0 & v_{ij}%
\end{array}%
\right)
\end{equation}%
with respect to the standard basis vectors $\left\vert \uparrow
\uparrow
\right\rangle $, $\left\vert \uparrow \downarrow \right\rangle $, $%
|\downarrow \uparrow \rangle $, and $\left\vert \downarrow
\downarrow \right\rangle $. Correspondingly the concurrence can be
calculated by
\begin{equation*}
C_{ij}=\max \left\{ 0,2\left( \left\vert z_{ij}\right\vert
-v_{ij}\right) \right\} .
\end{equation*}%
Furthermore, for the groundstate with vanishing total spin, it can
be connected to the isotropic correlation function $\left\langle
\mathbf{\sigma }_{i}\cdot \mathbf{\sigma }_{j}\right\rangle $ by
\cite{Wang}
\begin{equation}
C_{ij}=\frac{1}{2}\max \left\{ 0,-\left\langle \mathbf{\sigma
}_{i}\cdot \mathbf{\sigma }_{j}\right\rangle -1\right\} .
\end{equation}%
Analytical and numerical results show that the ground states of
the Hamiltonian (1) for the sizes we concern have spin zero. Then
the concurrences can be obtained directly from the corresponding
correlation functions. Now we define the $\alpha $ type
concurrence
\begin{equation}
C^{[\alpha ]}=\sum_{i}^{N-1}C_{i,i+\alpha },
\end{equation}%
where $\alpha =1,2,...,N/2$, and the total concurrence as the
summation of all types concurrences, i.e.
\begin{equation}
C_{T}=\sum_{\alpha }^{\frac{N}{2}}C^{[\alpha ]}.
\end{equation}%
It is obvious that $C_{T}$ shares the same property as the average
concurrence \cite{Yang}. We also define the concurrence jump as
$\triangle ^{\lbrack \alpha ]}=|C_{L}^{[\alpha ]}-C_{R}^{[\alpha
]}|$ ($\triangle _{T}=|C_{TL}-C_{TR}|$) denoting the $\alpha $
(total) type concurrence difference between the two ground states
at left ($L:J=J_{c}-0^{+}$) and right ($R:J=J_{c}+0^{+}$) side of
the critical points.

The conjectures of the relationship between entanglement and QPTs
in Refs \cite{LAWu:04,Sachdev} tell us that the concurrence, a
measure of entanglement, should be changed largely at the critical
points. So we calculate concurrence at point $B$, for $N=8,10$ and
$12$ systems numerically. The corresponding concurrence jumps
$\triangle ^{\lbrack \alpha ]}$ and $\triangle _{T}$ are listed in
Table. 1.

\begin{table}[tbp]
\begin{center}
\begin{tabular}{cccccc}
\hline\hline & \ \ \ $N$ \ \ \  & \ \ \ $8$ \ \ \  & \ \ \ $10$ \
\ \  & $12$ \ \ \  &
\\ \hline
& $\alpha=1$ & 0.7660 & 1.2755 & 0.6228 &  \\
& $\alpha=2$ & 1.8228 & 0 & 0 &  \\
& $\alpha>2$ & 0 & 0 & 0 &  \\
& total & 1.0568 & 1.2755 & 0.6288 &  \\ \hline
\end{tabular}%
\end{center}
\caption{\textit{The jumps of the pairwise concurrences $C^{[\protect\alpha %
]}$ and total concurrence $C_{T}$ at the critical point $B$ for
the $N=8,10$ and $12$ systems. It implies that not all types of,
but total concurrence are appropriate to depict the QPTs.}}
\end{table}
The big concurrence jumps $\triangle ^{\lbrack 1]}$ for $N=8,10,12$ and $%
\triangle ^{\lbrack 2]}$ for $N=8$ match the energy-level crossing
in Fig. 1, while the rest concurrences show no special behavior at
point $B$. It indicates that, at the critical point, not all types
of concurrence change largely. It shows that different types of
concurrences show different behaviors around the critical points.
It seems that there exists no single preferable type of
concurrence in characterizing the QPTs. A natural way to treat
this problem is to use the summation of all types of concurrence.
In
the following section, we will investigate the critical behavior at point $A$%
. The results further demonstrate that the total concurrence
$C_{T}$ seems to be a good candidate to depict the QPTs.

\section{III. Momentum Jump in QPTs Characterized by Entanglement}

According to quantum mechanics, we can always find out a
conservative quantity to distinguish the two degenerate ground
states at the energy-level crossing point. In this section we
study the critical behaviors at the energy-level crossing point
$A$ and $B$ (Fig. 2). We will show that these critical points are
just between the phases with momenta $0$ and $\pi$.

In the case $J=J_{0}/2$, the exact ground states of the $N$ site
system can be explicitly expressed as
\begin{equation}
\left\vert \phi _{1}\right\rangle =\left[ 12\right] \left[
34\right] ..\left[ N-1\text{ }N\right]
\end{equation}%
or
\begin{equation}
\left\vert \phi _{2}\right\rangle =\left[ 23\right] \left[
45\right] ..\left[ N1\right]
\end{equation}%
which are the direct products of the resonant valence bond (RVB) states $%
[ij]=(\left\vert \uparrow \downarrow \right\rangle
-\left\vert\downarrow
\uparrow \right\rangle)/\sqrt{2}$ of two spins located at the lattice sites $%
i$ and $j$ \cite{RVB}. Obversely, $\left\vert \phi _{1}\right\rangle $ and $%
\left\vert \phi _{2}\right\rangle $ are not orthogonal except in
the case of thermodynamical limit, but their combinations
$\left\vert \phi _{1}\right\rangle -\left\vert \phi
_{2}\right\rangle $ and $\left\vert \phi _{1}\right\rangle
+\left\vert \phi _{2}\right\rangle $ are two orthogonal degenerate
ground states.

Because the Hamiltonian is invariant under the translational transformation $%
T$, where $T\left\vert \uparrow \right\rangle _{i}=\left\vert
\uparrow \right\rangle _{i+1}$, the common eigenstates $\left\vert
\psi \right\rangle
=\left\vert \psi (\mathbf{S}_{1},\mathbf{S}_{2,}...,\mathbf{S}%
_{N})\right\rangle $ of $H$ and $T$ has momentum%
\begin{equation*}
k=\frac{2\pi n}{N}\equiv na,n=1,2,...,N
\end{equation*}%
which satisfies $T\left\vert \psi \right\rangle =\exp
(ina)\left\vert \psi \right\rangle $. Actually as mentioned above,
at point $J=J_{0}/2$, one can construct the two degenerate ground
states as
\begin{eqnarray}
\left\vert \psi _{1}\right\rangle  &=&\frac{1}{\sqrt{\Omega
_{1}}}\left[ \left\vert \phi _{1}\right\rangle -\left\vert \phi
_{2}\right\rangle \right]
,  \notag \\
\left\vert \psi _{2}\right\rangle  &=&\frac{1}{\sqrt{\Omega
_{2}}}\left[ \left\vert \phi _{1}\right\rangle +\left\vert \phi
_{2}\right\rangle \right] ,  \label{state}
\end{eqnarray}%
with momentum $k=0$ and $\pi $ respectively. Here%
\begin{eqnarray*}
\Omega _{1} &=&\langle \phi _{1}\left\vert \phi _{1}\right\rangle
+\langle \phi _{2}\left\vert \phi _{2}\right\rangle -2Re(\langle
\phi _{1}\left\vert
\phi _{2}\right\rangle ), \\
\Omega _{2} &=&\langle \phi _{1}\left\vert \phi _{1}\right\rangle
+\langle \phi _{2}\left\vert \phi _{2}\right\rangle +2Re(\langle
\phi _{1}\left\vert \phi _{2}\right\rangle ),
\end{eqnarray*}%
are the normalization factors.

We take a system of small size as an analytical illustration for
the momentum jump in QPT. For the system of $N=6$, the ground
states of the Hamiltonian can be obtained exactly in the whole
range of $J$ as
\begin{equation}
\left\vert \psi _{g}\right\rangle =\left\{
\begin{array}{ll}
\left\vert \psi _{1}\right\rangle & (J\geq J_{0}/2) \\
(\left\vert \psi _{2}\right\rangle +\eta \left\vert \psi _{e}\right\rangle )/%
\sqrt{\Omega _{3}} & (J\leq J_{0}/2)%
\end{array}%
\right.
\end{equation}%
corresponding to the eigenvalues

\begin{eqnarray}
E_{1} &=&-3(J_{0}+J)/2,  \notag \\
E_{2} &=&\left( \eta -5/2\right) J_{0}+\left( 1/2-\eta \right) J,
\end{eqnarray}%
respectively. Here
\begin{eqnarray}
\Omega _{3} &=&8\eta ^{2}-8\eta +20,  \notag \\
\eta &=&\frac{J-3J_{0}+\sqrt{9J^{2}-18JJ_{0}+13J_{0}^{2}}}{2\left(
J-J_{0}\right) },
\end{eqnarray}%
and
\begin{eqnarray}
\left\vert \psi _{e}\right\rangle &=&\frac{1}{2\sqrt{2}}%
[(1-T+T^{2}-T^{3}+T^{4}-T^{5})  \notag \\
&&\left\vert \uparrow \uparrow \uparrow \downarrow \downarrow
\downarrow \right\rangle -(1-T)\left\vert \uparrow \downarrow
\uparrow \downarrow \uparrow \downarrow \right\rangle )],
\end{eqnarray}%
is the excited state of the Hamiltonian (1) for $J=J_{0}$ with eigen energy $%
0$ and momentum $\pi $. It is easy to find that there is only one
critical point at which the energy-level crossing occurs. But for
$N>6$, there is one more energy-level crossing points at
$J>J_{0}/2$ as illustrated in Fig. 2.

Based on the exact results, the two spin concurrences are obtained
as
\begin{equation}
C^{[1]}=\left\{
\begin{array}{cc}
-4\left( \eta ^{2}+2\eta -2\right) /\Omega _{3} & \left( J<J_{0}/2\right) \\
0 & \left( J>J_{0}/2\right),%
\end{array}%
\right.
\end{equation}%
and
\begin{equation}
C^{[\alpha ]}=\left\{
\begin{array}{cc}
0 & \left( J<J_{0}/2\right) \\
0 & \left( J>J_{0}/2\right)%
\end{array}%
\text{ \ \ \ \ \ }(\alpha =2,3)\right. ,
\end{equation}%
which are plotted in Fig. 3.

\begin{figure}[tbp]
\includegraphics[bb=60 310 540 740, width=7 cm, clip]{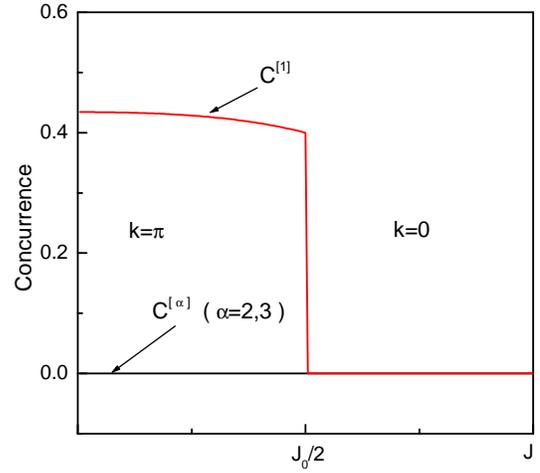}
\caption{\textit{Various types of concurrence as the function of $J$ for $%
N=6 $. The momenta of the ground state for $J<J_{0}/2$ and $J>J_{0}/2$ are $%
\protect\pi$ and $0$ respectively. It shows that only NN type of
concurrence possesses a jump around the critical point
$J=J_{0}/2$.}}
\end{figure}

It shows that the NN concurrence $C^{[1]}$ has a jump at the critical point $%
A$, while other types of concurrences $C^{[\alpha ]}$, $(\alpha
>1)$ do not have any special behavior. Obviously $C_{T}$ also has
a jump at the critical point, which is in agreement with the
observation presented in Sec. II. In the following section, we
will investigate the relations between various types of
concurrence jumps and a conservative quantity, momentum which has
a jump for the QPTs.

\section{IV. Concurrence jump in association with momentum jump in QPT}

From the above analysis, we know that the jumps of concurrences
$C^{[\alpha ]}$ at the critical points may be induced by the
energy-level crossing or the discontinuity of the groundstate
energy as a function of the coupling constants. On the other hand,
at the energy-level crossing point, the ground states are
degenerate. Thus an arbitrary linear combination of two degenerate
ground states is also the ground state. If a certain type of
concurrence has a jump at the critical point, the corresponding
concurrence of the combined ground state should be uncertain.
Meanwhile the difference of the concurrences between the two
orthogonal combined ground states should also depend on the way of
the combination. On the other hand, as the energy-level crossing
there must exist a conservative quantity which also experience a
jump. Then the phase separation can be also well described in
association with the jump of such a quantity. In this paper, this
conservative quantity is \emph{momentum} which is the generator of
translation. In general, one may say that the discontinuity of
$\partial E_{g}/\partial J$ leads to the jump of concurrence at
the critical points, but on the other hand, one can also say that,
it is the discontinuity of \textit{momentum} of the ground state
that leads to the jump of concurrence at the critical point.

In order to investigate the role that the momentum plays on the
change of various types of concurrence $C^{[\alpha ]}$, we
reconstruct two degenerate ground states at the critical point $A$
as
\begin{eqnarray}
\left\vert \Psi _{1}\right\rangle  &=&\cos \frac{\theta
}{2}\left\vert \psi _{1}\right\rangle +e^{i\phi }\sin \frac{\theta
}{2}\left\vert \psi
_{2}\right\rangle ,  \notag \\
\left\vert \Psi _{2}\right\rangle  &=&e^{-i\phi }\sin \frac{\theta }{2}%
\left\vert \psi _{1}\right\rangle -\cos \frac{\theta
}{2}\left\vert \psi _{2}\right\rangle ,
\end{eqnarray}%
where $\theta \in \left[ 0,\pi \right] $ and $\phi \in \left[
0,2\pi \right]
$. For $N$-site systems ($N\geq 6$), the pairwise concurrences of type $%
\alpha $ (total concurrence) of the two degenerate ground states
$\left\vert \Psi _{1}\right\rangle $ and $\left\vert \Psi
_{2}\right\rangle $ are
denoted as $C_{1}^{[\alpha ]}$ and $C_{2}^{[\alpha ]}$ ($C_{T1}$ and $C_{T2}$%
), respectively. A straightforward calculation shows that
$C_{1}^{[\alpha ]}=C_{2}^{[\alpha ]}=0$ for $\alpha >1$, while
$C_{1}^{[1]}$ and $C_{2}^{[1]} $ are nonzero and depend on the
parameters $\theta $ and $\phi $. Obviously, states $\left\vert
\Psi _{1}\right\rangle $ and $\left\vert \Psi _{2}\right\rangle $
are no more the eigenstates of momentum in the general case.

What we concern is the role the momentum plays on the jump of the
concurrence. For $N>6$ the difference between $C_{1}^{[1]}$ and
$C_{2}^{[1]}$ is
\begin{equation}
|C_{1}^{[1]}-C_{2}^{[1]}|=\left\vert \varepsilon _{11}+\varepsilon
_{12}-\varepsilon _{21}-\varepsilon _{22}\right\vert  \label{cc}
\end{equation}%
where%
\begin{equation}
\varepsilon _{ij}=\frac{1}{4}H\left( 3G_{ij}-1\right) \left(
3Q_{ij}-1\right) ,
\end{equation}%
\begin{eqnarray}
Q_{ij} &=&1-2\chi _{N}^{2}+(-)^{i}\xi _{N}\chi _{N}^{2}\cos \theta  \notag \\
&&+(-)^{i+j}\chi _{N}\cos \phi \sin \theta ,
\end{eqnarray}%
\begin{equation}
\chi _{N}\equiv \left[ 4-\xi _{N}^{2}\right] ^{-\frac{1}{2}},\xi
_{N}\equiv \left( \frac{1}{2}\right) ^{N/2-2}
\end{equation}%
and
\begin{equation}
H(x)=\left\{
\begin{array}{c}
1\text{\ \ \ \ }(x>0) \\
0\text{\ \ \ \ }(x<0)%
\end{array}%
\right.
\end{equation}%
is the Heaviside step function. In the following, we will show
that this difference reaches maxima at $\theta =0,\pi $ for any
$\phi \in \left[ 0,2\pi \right] $.

Notice that $3Q_{ij}-1\geq 0$ always holds when $\phi =\pi /2,
3\pi /2$. Then Eq. (\ref{cc}) can be rewritten as
\begin{equation}
|C_{1}^{[1]}-C_{2}^{[1]}|=3\left\vert \xi _{N}\chi _{N}^{2}\cos
\theta \right\vert .  \label{maximum 1}
\end{equation}%
It is a monotonic function, which reaches its maxima at $\theta
=0, \pi$.

Now we prove that for any $\theta$, the inequality
\begin{equation}
|C_{1}^{[1]}-C_{2}^{[1]}|\leq 3\left\vert \xi _{N}\chi
_{N}^{2}\cos \theta \right\vert  \label{maximum 2}
\end{equation}%
holds in the all range of $\phi$. It is convenient to consider the
problem in the range $\theta, \phi \in \left[ 0, \pi /2\right]$
without loss of generality. Since the functions $\sin \theta, \cos
\theta$ and $\cos \phi $ are positive in this range, we can prove
the above conclusion in the following three cases:

Case 1:
\begin{eqnarray}
3Q_{12}-1 &=&A-3\chi _{N}\cos \phi \sin \theta >0,  \notag \\
3Q_{21}-1 &=&B-3\chi _{N}\cos \phi \sin \theta >0;  \label{proof
begin}
\end{eqnarray}

Case 2:
\begin{eqnarray}
3Q_{12}-1 &=&A-3\chi _{N}\cos \phi \sin \theta <0,  \notag \\
3Q_{21}-1 &=&B-3\chi _{N}\cos \phi \sin \theta >0;  \label{case 2}
\end{eqnarray}%
and

Case 3:
\begin{eqnarray}
3Q_{12}-1 &=&A-3\chi _{N}\cos \phi \sin \theta <0,  \notag \\
3Q_{21}-1 &=&B-3\chi _{N}\cos \phi \sin \theta <0;
\end{eqnarray}%
where we have defined
\begin{eqnarray}
A &=&2-6\chi _{N}^{2}-3\xi _{N}\chi _{N}^{2}\cos \theta >0,  \notag \\
B &=&2-6\chi _{N}^{2}+3\xi _{N}\chi _{N}^{2}\cos \theta >0, \\
B &\geq &A.  \notag
\end{eqnarray}%
For all the above three cases we have
\begin{eqnarray}
3Q_{11}-1 &=&A+3\chi _{N}\cos \phi \sin \theta >0,  \notag \\
3Q_{22}-1 &=&B+3\chi _{N}\cos \phi \sin \theta >0.
\end{eqnarray}
Actually, in case 1, we have
\begin{equation*}
|C_{1}^{[1]}-C_{2}^{[1]}|=3\left\vert \xi _{N}\chi _{N}^{2}\cos
\theta \right\vert .
\end{equation*}%
In case 2, from Eq.(\ref{case 2}) we have
\begin{equation}
\left\vert D\right\vert <\left\vert -4\xi _{N}\chi _{N}^{2}\cos
\theta \right\vert ,
\end{equation}%
where
\begin{eqnarray}
D &=&-\frac{2}{3}-3\xi _{N}\chi _{N}^{2}\cos \theta  \notag \\
&&+\chi _{N}\cos \phi \sin \theta +2\chi _{N}^{2}.
\end{eqnarray}%
Then the concurrence difference is
\begin{eqnarray}
|C_{1}^{[1]}-C_{2}^{[1]}| &=&\left\vert \varepsilon
_{11}-\varepsilon
_{21}-\varepsilon _{22}\right\vert  \notag \\
&=&\frac{3}{4}\left\vert D\right\vert <3\left\vert \xi _{N}\chi
_{N}^{2}\cos \theta \right\vert .
\end{eqnarray}%
Similarly, for the case 3, we have%
\begin{equation}
|C_{1}^{[1]}-C_{2}^{[1]}|=\frac{3}{2}\left\vert \xi _{N}\chi
_{N}^{2}\cos \theta \right\vert <3\left\vert \xi _{N}\chi
_{N}^{2}\cos \theta \right\vert .  \label{proof end}
\end{equation}
For the cases that $\theta $ and $\phi $ are taken in the rest
ranges, similar proof as presented in (\ref{proof
begin})--(\ref{proof end}) can get the same conclusion. So from
Eqs. (\ref{maximum 1}) and (\ref{maximum 2}) we
conclude that the concurrence difference between $C_{1}^{[1]}$ and $%
C_{2}^{[1]}$ reaches the maxima when $\theta =0, \pi$ for any
$\phi $.

In fact, for the cases $N\geq 8$ within the special domain of
$\theta \sim 0,\pi $ and any $\phi $, it always hold that
$3Q_{ij}-1>0$. Hence we have
\begin{equation}
|C_{1}^{[1]}-C_{2}^{[1]}|=\left\{
\begin{array}{ccc}
&3\xi _{N}\chi _{N}^{2}(1-\theta
^{2})\text{\ \ }&(\theta \sim 0) \\
&3\xi _{N}\chi _{N}^{2}[1-(\pi -\theta )^{2}]\text{\ \ }&(\theta \sim \pi)%
\end{array}%
\right.
\end{equation}
which reaches its maxima $3\xi _{N}\chi _{N}^{2}$ at $\theta
=0,\pi $. Here the two combined states $\left\vert \Psi
_{1}\right\rangle $ and $\left\vert \Psi _{2}\right\rangle $ are
just the eigenstates of momentum, i.e. the ground states at
$J=J_{c}\pm 0^{+}$.

As illustration, in Fig. 4 the concurrence differences,
$|C_{1}^{[\alpha
]}-C_{2}^{[\alpha ]}|$ as the functions of the parameters $\theta $ and $%
\phi $ are plotted for $N=6,8,12$ and $16$ systems.

\begin{figure}[tbp]
\includegraphics[bb=20 230 580 740, width=8 cm, clip]{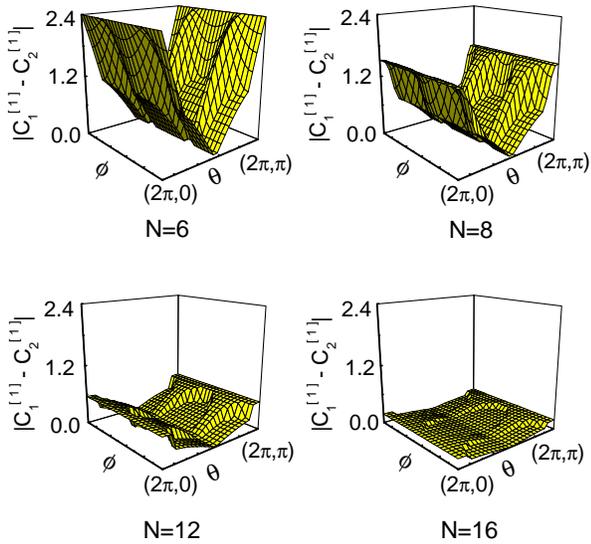}
\caption{\textit{The NN concurrence difference of the two
reconstructed degenerate ground states for $N=6,8,12$ and $16$.
Since all the rest types
of concurrences are $0$, the behavior of $|C_{1}^{[1]}-C_{2}^{[1]}|$ and $%
|C_{T1}-C_{T2}|$ are the same. It shows that the total concurrence
difference reach the maxima when $\protect\theta =0$ or
$\protect\pi $ for all $\protect\phi $ and $N$. But the maxima
decay exponentially with the size $N$ of the system.}}
\end{figure}
It shows that there always exists a maximal concurrence
difference, or a maximal concurrence jump, at the points where the
two degenerate ground states are the eigenstates of momentum with
$k=0,\pi $. Furthermore, in the
thermodynamic limit, we have $\lim_{N\rightarrow \infty }\xi _{N}=0$ and $%
\lim_{N\rightarrow \infty }\chi _{N}=\frac{1}{2}$, and then the
maxima decays to zero exponentially. The results indicate that the
change of the
momentum induces the maximal jump of $|C^{[1]}|$ or $|C_{T}|$ at the point $%
A $. So far, we can not say which types of concurrence is
preferable to characterize the QPT, from the following arguments,
we will see that an anomalous result for point $B$ will give a
final selection.

Now we turn to consider the situation at the critical point $B$.
In this case, the differences of concurrence $|C_{1}^{[\alpha
]}-C_{2}^{[\alpha ]}|$ and $|C_{T1}-C_{T2}|$ can not be obtained
analytically. The numerical method is employed to calculate the
differences of various types of concurrences.
All types of concurrence difference as the functions of the parameters $%
\theta$ and $\phi$ are plotted in Fig. 5.

\begin{figure}[tbp]
\includegraphics[bb=20 215 580 760, width=8 cm, clip]{fig5(a).eps} %
\includegraphics[bb=20 210 580 760, width=8 cm, clip]{fig5(b).eps}
\caption{\textit{All types of concurrence differences of the
degenerate ground states for $N=8$ (a) and $10,12$ (b). Notice
that the anomalous behavior in $N=8$ case shows that only the
total concurrence difference
reach the maxima when $\protect\theta =0$ or $\protect\pi $ for all $\protect%
\phi $ and $N$.}}
\end{figure}
In Fig. 5(a), it shows that the pairwise concurrences of NN and
NNN for $N=8$ have jumps, while the rest have no jumps. This
result is different from that for the $A$ point, in which case
there is no jump for NNN concurrence. Another interesting result
is that the difference $|C_{1}^{[1]}-C_{2}^{[1]}|$ does not reach
the maxima when the corresponding two degenerate states are the
eigenstates of momentum. This anomalous phenomenon indicates that
the NN concurrence seems not to be sufficient characterize the
QPT. However, Fig. 5(a) also shows that the difference
$|C_{T1}-C_{T2}|$ still obeys the same
rule we obtained at point $A$. In Fig. 5(b) the corresponding results for $%
N=10$ and $12$ are plotted, which are similar to that of point
$A$. Thus all the results imply that the difference of total
concurrence reach the maxima when two degenerate states are the
eigenstates of momentum. In other words, it is the change of
momentum that induces the jump of total concurrence. Based on all
the results, we conclude that the total concurrence is a good
candidate to characterize the quantum critical behavior for the
frustrated spin ring systems we concerned.

\section{V. Discussion}

Summing up, in this paper we have shown how to establish the
connection between the concurrence jump and the change of
groundstate momentum at the QPT critical point. All types of
pairwise concurrence are investigated
analytically and numerically. The results for both critical points $A$ and $%
B $ indicate that the difference of total concurrence reach the
maxima when the two degenerate groundstates are just the
eigenstates of momentum. It also reveal another interesting
relation between correlation function and the concurrence. As
mentioned in section II, the NN and NNN correlation function must
have a jump at the energy-level crossing points, while the NN and
NNN concurrences may not. But when the total concurrence is
considered, it must have a jump at the critical points. However,
the total correlation
function $\sum_{ij}\langle \mathbf{S}_{i}\cdot \mathbf{S}_{j}\rangle _{g}=%
\frac{1}{2}\langle
(\mathbf{S}^{2}-\sum_{i}\mathbf{S}_{i}^{2})\rangle _{g}$ (where
$\mathbf{S}=\sum_{i}\mathbf{S}_{i}$ is the total spin) has no jump
at the critical points since the spins of the ground states are
zero.

Concerning the model for the case $J_{1}\neq J_{2}$ as illustrated
in Fig. 1, a straight forward calculation shows that the states
$\left\vert \psi
_{1}\right\rangle $ and $\left\vert \psi _{2}\right\rangle $ defined in (\ref%
{state}) are still the degenerate groundstates of the Hamiltonian if $%
J_{1}+J_{2}=J_{0}$. This means that $J_{1}+J_{2}=J_{0}$ is the
boundary of different quantum phases in the $J_{1}-J_{2}$ plane.
Starting from this observation we can extend our study to the more
general case of $J_{1}\neq J_{2}$ to verify the conclusion
obtained in this paper. It will appear in a successive paper.

We acknowledge the support of the CNSF (grant No. 90203018,
10474104), the Knowledge Innovation Program (KIP) of Chinese
Academy of Sciences, the National Fundamental Research Program of
China (No. 001GB309310).

\end{document}